\documentclass[aps,prl,showpacs,twocolumn,superscriptaddress]{revtex4}
\usepackage{graphicx}
\usepackage{wrapfig}
\usepackage{multirow}

\begin{document}

%\preprint{}

\title{Mechanisms in knockout reactions}


\author{D. Bazin}
\email[]{bazin@nscl.msu.edu}
\affiliation{National Superconducting Cyclotron Laboratory, Michigan State University, East Lansing, MI 48824, USA}

\author{R. J. Charity}
\affiliation{Department of Chemistry, Washington University, St. Louis, MO 63130, USA}

\author{R. T. de Souza}
\affiliation{Department of Chemistry, Indiana University, Bloomington, IN 47405, USA}

\author{M. A. Famiano}
\affiliation{Department of Physics, Western Michigan University, Kalamazoo, MI 49008, USA}
\affiliation{National Superconducting Cyclotron Laboratory, Michigan State University, East Lansing, MI 48824, USA}

\author{A. Gade}
\affiliation{National Superconducting Cyclotron Laboratory, Michigan State University, East Lansing, MI 48824, USA}
\affiliation{Department of Physics and Astronomy, Michigan State University, East Lansing, MI 48824, USA}

\author{V. Henzl}
\affiliation{National Superconducting Cyclotron Laboratory, Michigan State University, East Lansing, MI 48824, USA}

\author{D. Henzlova}
\affiliation{National Superconducting Cyclotron Laboratory, Michigan State University, East Lansing, MI 48824, USA}

\author{S. Hudan}
\affiliation{Department of Chemistry, Indiana University, Bloomington, IN 47405, USA}

\author{J. Lee}
\affiliation{National Superconducting Cyclotron Laboratory, Michigan State University, East Lansing, MI 48824, USA}

\author{S. Lukyanov}
\affiliation{FLNR/JINR, 141980 Dubna, Moscow region, Russia}
\affiliation{National Superconducting Cyclotron Laboratory, Michigan State University, East Lansing, MI 48824, USA}

\author{W. G. Lynch}
\affiliation{National Superconducting Cyclotron Laboratory, Michigan State University, East Lansing, MI 48824, USA}
\affiliation{Department of Physics and Astronomy, Michigan State University, East Lansing, MI 48824, USA}

\author{S. McDaniel}
\affiliation{National Superconducting Cyclotron Laboratory, Michigan State University, East Lansing, MI 48824, USA}

\author{M. Mocko}
\affiliation{Los Alamos National Laboratory, Los Alamos NM 87545, USA}
\affiliation{National Superconducting Cyclotron Laboratory, Michigan State University, East Lansing, MI 48824, USA}

\author{A. Obertelli}
\affiliation{DAPNIA/SPhN, CEA Saclay, F-91191 Gif-sur-Yvette Cedex, France}
\affiliation{National Superconducting Cyclotron Laboratory, Michigan State University, East Lansing, MI 48824, USA}

\author{A. M. Rogers}
\affiliation{National Superconducting Cyclotron Laboratory, Michigan State University, East Lansing, MI 48824, USA}

\author{L. G. Sobotka}
\affiliation{Department of Chemistry, Washington University, St. Louis, MO 63130, USA}

\author{J. R. Terry}
\affiliation{Department of Physics, Yale University,  New Haven, CT 06520, USA}
\affiliation{National Superconducting Cyclotron Laboratory, Michigan State University, East Lansing, MI 48824, USA}

\author{J. A. Tostevin}
\affiliation{Department of Physics, University of Surrey, Guildford GU2 7XH, UK}
\affiliation{National Superconducting Cyclotron Laboratory, Michigan State University, East Lansing, MI 48824, USA}

\author{M. B. Tsang}
\affiliation{National Superconducting Cyclotron Laboratory, Michigan State University, East Lansing, MI 48824, USA}

\author{M. S. Wallace}
\affiliation{Los Alamos National Laboratory, Los Alamos NM 87545, USA}
\affiliation{National Superconducting Cyclotron Laboratory, Michigan State University, East Lansing, MI 48824, USA}

\noaffiliation

\date{\today}

\begin{abstract}
We report on the first detailed study of the mechanisms involved in knockout reactions, via a coincidence measurement of the residue and fast proton in one-proton knockout reactions, using the S800 spectrograph in combination with the HiRA detector array at the NSCL.
Results on the reactions $^9$Be($^9$C,$^8$B+X)Y and $^9$Be($^8$B,$^7$Be+X)Y are presented. 
They are compared with theoretical predictions for both the diffraction and stripping reaction mechanisms, as calculated in the eikonal model.  
The data shows a clear distinction between the two reaction mechanisms, and the observed respective proportions are very well reproduced by the reaction theory.  
This agreement supports the results of knockout reaction analyses and their applications to the spectroscopy of rare isotopes.
\end{abstract}

%\keywords{}

\maketitle

%\section{Introduction}
In a nucleon knockout reaction, a single nucleon is removed from a fast moving projectile in the high energy collision with a light target.
This type of reaction is being used extensively to probe the wave functions of rare isotopes produced at fragmentation facilities~\cite{knockout}.  
The success of this technique relies on the utilization of fast radioactive beams, which on one hand provide a high luminosity due to the use of thick targets, and on the other hand enables simple assumptions such as the sudden and eikonal approximations in the reaction theory.  
This unique combination has already yielded a vast amount of spectroscopic information on nuclei far from stability, where benchmarking nuclear models such as the shell model is crucial.

Typical experiments employ thick targets and $\gamma$-ray spectroscopy for luminosity, and the removed nucleon is not detected.
Since, therefore, no information is recorded experimentally on the fate of the removed nucleon or the final state of the target nucleus, the reaction theory is used to estimate both elastic (also called diffraction) and inelastic (also called stripping) breakup mechanisms.  
These contributions to the knockout cross section are summed before comparison with experiment.  
It is therefore of great importance to investigate the proportion of each reaction mechanism  experimentally in order to validate the reaction theory, and assess any deviations due to approximations in the reaction dynamics.  
In particular, no experimental data showing the kinematics differences between the two mechanisms has been available so far.

A more exclusive experiment where the momentum vectors of both the heavy residue and removed nucleon are measured is necessary to distinguish between the different reaction mechanisms. 
The one-proton knockout reactions on $^9$C and $^8$B were chosen for several reasons, summarized in table \ref{Eikonal}.  
The diffraction component in the proton knockout from $^8$B should be enhanced in comparison to $^9$C due to its lower proton separation energy.  
The final states of both heavy residues are well defined since $^8$B has no bound excited state and $^7$Be has only one.  
Finally, the possibility of producing a radioactive beam containing both nuclei, as well as the easier detection of the knocked out protons rather than neutrons, were further motivations.  
The eikonal calculations were performed using proton bound state and target and residue density parameters taken from previous studies \cite{BHST}.  
The Coulomb breakup cross sections are also included since both contribute to the diffractive component.

\begin{table}
\caption{\label{Eikonal}The stripping and diffraction components of the one-proton knockout cross section for $^9$C and $^8$B are calculated in the eikonal model.  
The small contributions from Coulomb breakup are taken from \cite{Enders}, which used very similar incident energies.  
The spectroscopic factors C$^2$S$_{SM}$ are taken from shell-model calculations using the PJT interaction~\cite{SM}.  
The radial mismatch factor M for the $^9$C knockout comes from the imperfect overlap of the least bound proton wave functions between the projectile and the residue, as calculated in \cite{Enders}. 
The last column indicates the predicted proportions of elastic breakup or diffraction (including Coulomb) relative to the total.}
 \begin{ruledtabular}
 \begin{tabular}{ccccccccc}
Initial&Final&S$_p$&$\sigma_{str}$&$\sigma_{diff}$&$\sigma_{C}$&C$^2$S$_{SM}$&M&\%$_{diff}$\\
state & state & (MeV) & (mb) & (mb) & & & \\ \hline
$^9$C&$^8$B&1.296&44.57&15.27&1.1&0.94 & 0.976 & 26.9 \\
$^8$B&$^7$Be$_{ gs}$&0.137&64.42&31.65&7.7&1.036&1&\multirow{2}{*}{37.1}\\
$^8$B&$^7$Be$_{ex}$&0.566&57.34&24.44&3.4&0.22&1&\\
% Lines of table here ending with \\
 \end{tabular}
 \end{ruledtabular}
 \end{table}

The experiment was carried out at the National Superconducting Cyclotron Laboratory (NSCL), where a primary beam of $^{16}$O at 150 MeV/u was used to produce the $^9$C and $^8$B radioactive beams via projectile fragmentation on a 1763 mg/cm$^2$ $^9$Be target.  
Both $^9$C and $^8$B beams were filtered out simultaneously by the A1900 fragment separator~\cite{A1900}, and contained a small amount of $^7$Be and $^6$Li contamination.  
Reactions from the different components of this cocktail beam could be easily identified event-by-event, using the time-of-flight measured between two plastic scintillators located in the beam line before the reaction target.  
The one-proton knockout reactions took place in the scattering chamber of the S800 spectrograph~\cite{S800}, on a 188 mg/cm$^2$ $^9$Be target surrounded by the HiRA detector array~\cite{HiRA}.   
The mid-target energies of the $^9$C and $^8$B beams were 97.9 MeV/u and 86.7 MeV/u respectively.  
The one-proton knockout residues were collected and identified around 0$^{\circ}$ degree by the S800 spectrograph, whereas the light particles emerging at large angles were detected and identified by the HiRA array.  
The angles covered ranged from 11$^{\circ}$ to 60$^{\circ}$.  
Due to the finite 5\% momentum acceptance of the S800 spectrograph, several overlapping magnetic rigidity settings were necessary to cover the range of momentum spanned by the $^8$B and $^7$Be residues from both reactions.

The light-particle identification in the HiRA telescopes was performed using the signals from the silicon detector for energy loss versus the CsI crystal for total energy.  
The punch-through energy for protons in the CsI detector is around 120 MeV.  
From both reactions, events in coincidence with the one-proton knockout residues revealed not only protons, but deuterons and also a few tritons.  
Such events must involve an inelastic interaction with the target.  
For events in coincidence with protons only, however, both elastic and inelastic interactions are possible, corresponding to the diffraction and stripping mechanisms.  
To investigate the differences between these mechanisms experimentally, the energy of the detected proton was plotted versus that of the heavy residue, after proper calibrations of both the S800 spectrograph and the HiRA array.  
The energy of the heavy residue is reconstructed from the focal-plane positions and angles measurements via an inverse map calculated with the ion optics program COSY Infinity~\cite{COSY}.
\begin{figure}
\centering
\includegraphics[scale=0.65]{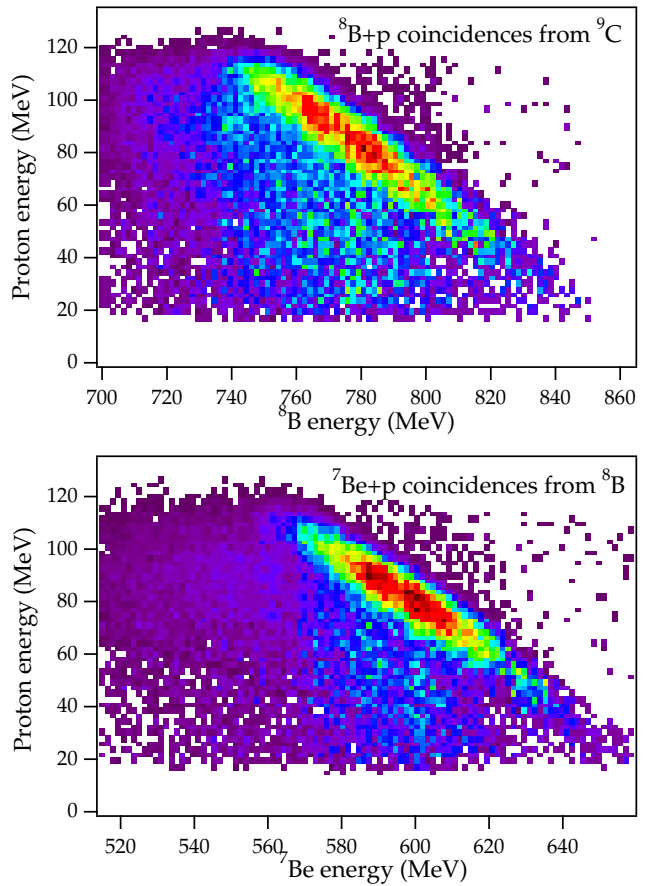}
\caption{(color online) Two dimensional spectra of the energy of protons and of the heavy residue in one-proton knockout reactions from $^9$C (top) and $^8$B (bottom) projectiles respectively.  
The narrow bands of constant energy sum correspond to elastic breakup whereas other events are associated with inelastic breakup (see text).}
\label{conserv}
\end{figure}
The resulting spectra are shown in Fig. \ref{conserv}, where a narrow band at high energy can be observed for both reactions.  
This narrow band corresponds to events in which there is minimal transfer of energy to the target and the incoming projectile kinetic energy is essentially shared between the heavy residue and the proton.  
These events are classified as due to elastic breakup.  
In contrast, the events located below the band correspond to reactions where a large portion of the initial kinetic energy of the projectile is lost to the target nucleons.  
These events are identified as inelastic breakup since they involve excitation of the target.

\begin{figure}
\centering
\includegraphics[scale=0.65]{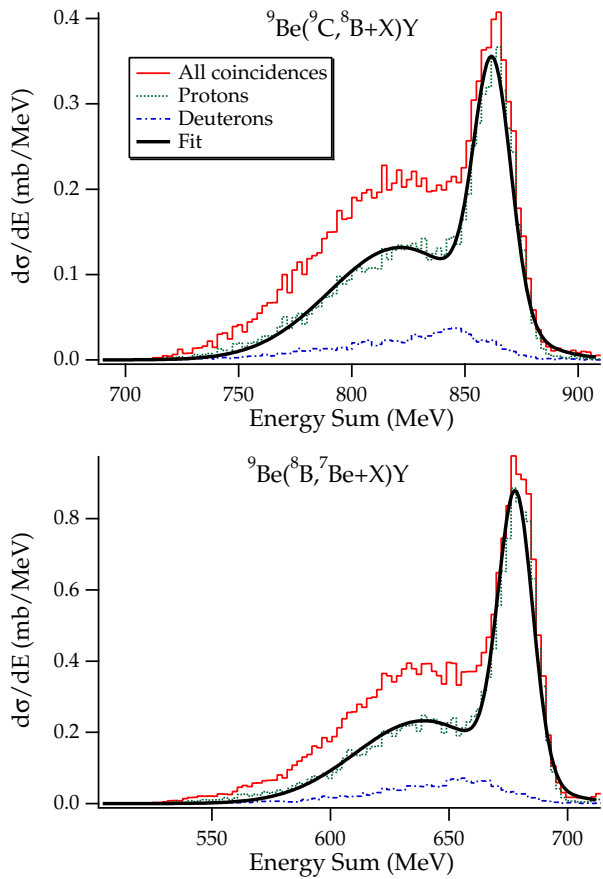}
\caption{(color online) Energy sum spectra of the one-proton knockout residue and the light particles detected in coincidence in the HiRA detector array for $^9$C (top) and $^8$B (bottom) projectiles.  The sharp peak corresponding to elastic breakup is visible in proton coincidence events, whereas it disappears for deuteron and other coincidence events (see text).}
\label{sum}
\end{figure}

To further characterize the two classes of events observed in the previous spectra, the energy sum spectra of the heavy residue and light particle detected in coincidence was reconstructed from individual energies.  
They are shown in Fig. \ref{sum} for all coincidence events, as well as for protons only and deuterons only.  
The events in coincidence with protons show two distinct features: a sharp peak at high energy and a broad, overlapping peak at lower energy.  
For events in coincidence with deuterons on the other hand, only the broad peak is visible.  
Since the neutron in the deuteron can only originate from the target, this further indicates that the sharp peak corresponds to elastic breakup processes, where the target stays in its ground state.  
The width of the elastic peak comes from both the momentum width of the incoming beam (1\%), and the broadening due to the differential energy loss in the target - the energy difference between reactions happening at the front and the back of the target.

In order to evaluate the proportion of elastic breakup in the reactions, the elastic cross sections have to be extracted from the data.  
The following procedure was used.
The scattering angle distributions of protons detected in HiRA for both inelastic and elastic peaks were extracted by applying a cut in the energy sum spectra at the junction between the two peaks.  
The elastic distributions were then obtained by subtracting the tail of the inelastic contamination leaking into the elastic peak above the junction, as determined from a double-Gaussian fit of the distributions (see Fig. \ref{sum}).  
The resulting proton scattering angle distributions were then corrected for the geometrical acceptance of the HiRA array within its angular coverage, obtained from a Monte-Carlo simulation.  
\begin{figure}
\centering
\includegraphics[scale=0.55]{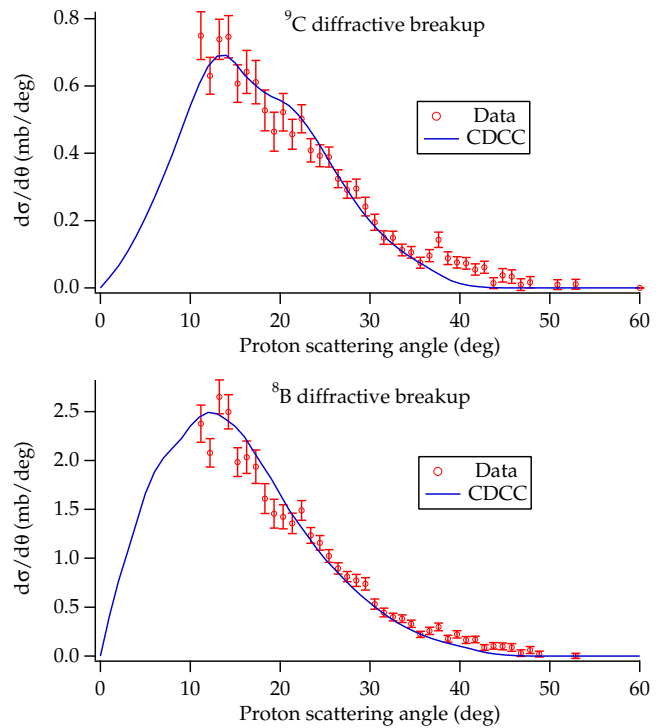}
\caption{Proton scattering angle distributions obtained for $^9$C (top) and $^8$B (bottom) elastic events, after subtraction of the inelastic contamination using the fitted energy sum spectra (see Fig. \ref{sum}).  
Also shown are the results from a CDCC calculation of the same quantity.  
The CDCC distributions are used to deduce the amount of unobserved cross section due to the lack of angular coverage between 0$^{\circ}$ and 10$^{\circ}$, after (minor) normalization to the data.}
\label{pscatter}
\end{figure}
The proton angular distributions obtained from the $^9$C and $^8$B elastic breakup events are shown in Fig. \ref{pscatter}. 
There they are compared with the theoretical predictions from Continuum Discretized Coupled Channels (CDCC) calculations, that retain the full three-body final state kinematics of the target, residue ($r$) and the diffracted proton. 
The CDCC calculations make use of the methodology of Ref.~\cite{Tos01} to calculate the laboratory frame multi-differential cross sections $d^3 \sigma / d\Omega_r d\Omega_p dE_p$ of the proton and $^8$B/$^7$Be residues that are then integrated over the angular acceptance ($\Delta \Omega_r$= 21 msr) of the fast, forward-going residue and all proton energies $E_p\leq 120$ MeV. 
The parameters used in the CDCC calculations are the same as employed for the earlier eikonal model results, including the complex proton-target and residue-target distorting potentials that were taken as the double- and single-folded interactions used to generate the corresponding eikonal elastic S-matrices.
The unobserved cross section between 0$^{\circ}$ and 10$^{\circ}$ could be inferred from the excellent agreement between the CDCC theory and the observed distributions at larger angles.  
Corrections of 15(3)\% and 28(5)\% were calculated for the $^9$C and $^8$B elastic breakup cross sections respectively, using the CDCC distributions.  
After combining the data from the various magnetic rigidity settings necessary to cover the momentum distributions of the heavy residues, correcting for detector efficiencies and the finite angular acceptance of the S800, the integrated proton scattering angle distributions yielded cross sections of 13.8(6) mb and 49(2) mb for the elastic breakup cross sections of $^9$C and $^8$B respectively.

\begin{figure}
\centering
\includegraphics[scale=0.55]{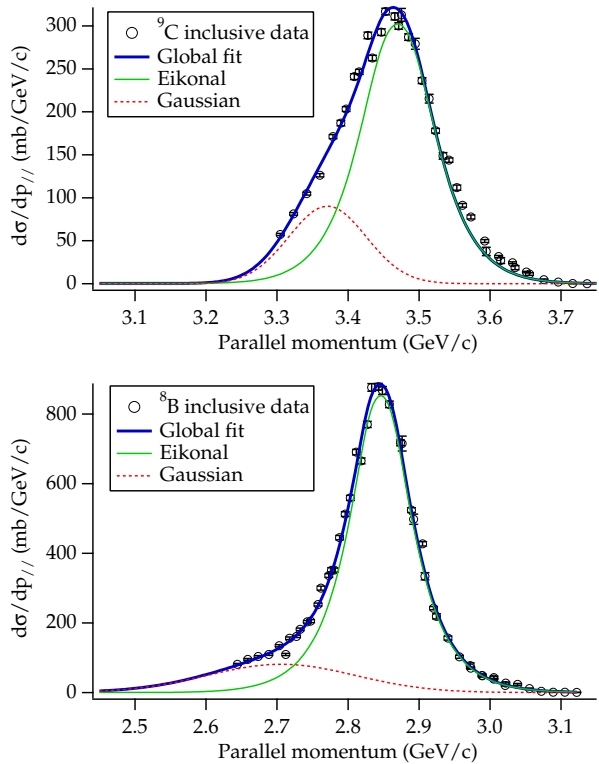}
\caption{(color online) Inclusive parallel momentum distributions of the heavy residues from the reactions $^9$Be($^9$C,$^8$B+X)Y (top) and $^9$Be($^8$B,$^7$Be+X)Y (bottom), reconstructed from the various magnetic rigidity settings of the S800. 
The shapes of the distributions calculated from the eikonal model reproduce the data very well, except for the low momentum tails. For that reason, Gaussian distributions have been added in the fits used to extract the inclusive cross sections.}
\label{inclusive}
\end{figure}

The inclusive one-proton knockout cross sections were obtained in a similar way as the elastic breakup cross section, except that no coincidence with the HiRA array was required.  
Fig. \ref{inclusive} shows the combined data obtained from the various settings of the S800 for the parallel momentum distribution of the heavy residue, as well as the eikonal calculation and the fit used to determine the inclusive cross section.  
The observed distributions are fitted by combining the eikonal distribution with a Gaussian centered at lower momentum to account for the low momentum tail.  
Note that for the eikonal component only the position and height were varied. 
The values obtained are 56(3) mb and 127(5) mb for the $^9$Be($^9$C,$^8$B+X)Y and $^9$Be($^8$B,$^7$Be+X)Y reactions respectively.  
They are in good agreement with the earlier results of Enders et al. \cite{Enders}.

\begin{table}
\caption{\label{results}Measured and theoretical diffraction (including Coulomb) components of the one-proton knockout cross section for $^9$C and $^8$B.
The calculated diffraction component agrees very well with the observation in both reactions.
The large error bars in \cite{Enders} come from the inclusive method used in that experiment.
The theoretical inclusive cross sections are shown, which include center-of-mass correction factors A/(A-1).
The deduced reduction factors R$_S$=$\sigma_{exp}/\sigma_{th}$ are also shown, as well as previous measurements.}
\begin{ruledtabular}
\begin{tabular}{c|ccc|cccc}
Proj.&\%$_{diff}$&\%$_{diff}$&\%$_{diff}$&$\sigma_{th}$&R$_S$&R$_S$&R$_S$\\
&\footnotemark[1]&\footnotemark[2]&\cite{Enders}&mb&\footnotemark[1]&\cite{Enders}&\cite{BHST}\\ \hline
$^9$C&25(2)&26.9&26(10)&62.90&0.84(5)&0.82(6)&-\\
$^8$B&38(3)&37.1&28(14)&144.28&0.88(4)&0.86(7)&0.88(4)\\
% Lines of table here ending with \\
\end{tabular}
\end{ruledtabular}
\footnotetext[1]{This work}
\footnotetext[2]{Calculated (from Table \ref{Eikonal})}
\end{table}

Our experiment measures the components of elastic and inelastic breakup in one-proton knockout reactions.  
The kinematic differences between the two reaction mechanisms are apparent in the experimental data.  
Moreover, our analysis shows that the observed diffraction and stripping contributions are very well reproduced by the theory, as shown in Table \ref{results}.  
This adds considerable support to the use of the eikonal model as a quantitative tool, able, for example, to determine single-particle spectroscopic strengths in rare isotopes.

\begin{acknowledgments}
This work was supported by the National Science Foundation under Grant No. PHY-0606007, and the United Kingdom Science and Technology Facilities Council (STFC) under Grant No. EP/D003628.
\end{acknowledgments}



\begin{thebibliography}{00}

\bibitem{knockout}
P.G. Hansen and J.A. Tostevin,
Annu. Rev. Nucl. Part. Sci. {\bf 53} (2003) 219.

\bibitem{BHST}
B. A. Brown, P. G. Hansen, B. M. Sherrill and J. A. Tostevin,
Phys. Rev. C {\bf 65}, (2002) 061601(R)

\bibitem{A1900}
D. J. Morrissey {\it et al.},
Nucl. Instrum. Methods Phys. Res. B{\bf 204} (2003) 90.

\bibitem{S800}
D. Bazin {\it et al.},
Nucl. Instrum. Methods Phys. Res. B{\bf 204} (2003) 629.

\bibitem{HiRA}
M. S. Wallace {\it et al.},
Nucl. Instrum. Methods Phys. Res. A{\bf 583} (2007) 302.

\bibitem{Enders}
J. Enders {\it et al.},
Phys. Rev. C{\bf 67} (2003) 064301.

\bibitem{SM}
B. A. Brown {\it et al.},
Prog. Part. Nucl. Phys. {\bf 47}, 517 (2001).

\bibitem{COSY}
M. Berz, K. Joh, J. A. Nolen, B. M. Sherrill, and A. F. Zeller,
Phys. Rev. C {\bf 47} (1993) 537.

\bibitem{Tos01}
J. A. Tostevin, F. M. Nunes and I. J. Thompson, Phys. Rev. C {\bf 63},
024617 (2001).

\end{thebibliography}
\end{document}